\documentclass[11pt]{article}
\usepackage{authblk}
\usepackage{mathrsfs}
\usepackage{amssymb}
\usepackage{color}
\usepackage{slashbox}
\usepackage{amsthm}
\usepackage{amsmath}
\allowdisplaybreaks[4]
\usepackage{graphicx}
\usepackage{subfigure}

 \theoremstyle{definition}

 \numberwithin{equation}{section}

\def\u0{u_{0}}
\def\l2{l_{2}}

\def\E{{\mathbb E}}

\theoremstyle{definition}

\theoremstyle{remark}

\makeatletter

\newcommand{\Rmnum}[1]{\expandafter\@slowromancap\romannumeral #1@}
\makeatother

\begin{document}
\title{A weight-bounded importance sampling method for variance reduction}
\author[1]{Tengchao Yu}
\author[2]{Linjun Lu}
\author[3]{Jinglai Li}
\affil[1]{School of Mathematical Sciences and Institute of Natural Sciences, 
Shanghai Jiao Tong University, 800 Dongchuan Rd, Shanghai 200240, China.}
\affil[2]{Corresponding Author, School of Naval Architecture, Ocean and Civil Engineering, Shanghai Jiao Tong University, Shanghai 200240, China.
Email: {linjunlu@sjtu.edu.cn}}
\affil[3]{Corresponding Author, Department of Mathematical Sciences, University of Liverpool, Liverpool L69 7ZL, UK.}
\date{}

\maketitle
\begin{abstract}
Importance sampling (IS) is an important technique to reduce the estimation variance in Monte Carlo simulations. 
In many practical problems, however, the use of IS method may result in
unbounded variance, and thus fail to provide reliable estimates. 
To address the issue, we propose a method which can prevent the risk of unbounded variance;
the proposed method performs the standard IS for the integral of interest in a region only in which the IS weight is bounded 
and use the result as an approximation to the original integral. 
It can be verified that the resulting estimator has 
a finite variance. Moreover, we also provide a normality test based method to identify the region with bounded IS weight (termed as the safe region)
from the samples drawn from the standard IS distribution. 
With numerical examples, we demonstrate that the proposed method can yield rather reliable estimate when 
the standard IS fails, and it also outperforms the defensive IS, a popular method to prevent unbounded variance. 
\end{abstract}


\section{Introduction}
The Monte Carlo (MC) method~\cite{liu2008monte,robert2013monte}, from a mathematical point of view, is a technique to evaluate integrals or expectations by random sampling.
Since its invention,  the MC method has found vast applications in many fields of science and engineering, ranging from statistical physics~\cite{landau2014guide} to financial engineering~\cite{glasserman2013monte}.
A well-known issue in the standard MC method is that it suffers from a rather slow convergence: the variance of an MC estimator is proportional to $1/\sqrt{n}$ with $n$ being the number of samples, and as a result,
it may require a rather large number of samples to produce a reliable estimate in many practical problems. 
To this end, the technique of importance sampling (IS)~\cite{liu2008monte,robert2013monte} is often used to reduce the variance, and simply speaking,
the IS method draws samples from an alternative distribution (known as the IS distribution) instead of the original one, and then corrects for the biasing caused by using the altering the distribution 
by assigning appropriate weight to each sample.  
Designing IS distribution is the key in the implementation of the IS method, and a good IS distribution
can significantly improve the sampling efficiency.  
On the other hand, if the sampling distribution is not properly designed, the IS simulation will perform poorly and in some extreme cases, it may fail completely,
in the sense that it results in infinite estimator variance \cite{glasserman1997counterexamples}.    
In this case, the IS method may yield completely wrong estimates. 
Unfortunately, it is usually not possible to know in advance whether the chosen IS distribution is appropriate. 
To this end, it becomes a rather important task to develop methods that can prevent the infinite estimator variance of standard IS. 
To address the issue, a scheme called defensive IS (DIS) was proposed in \cite{hesterberg1995weighted}, where the basic idea is use a mixture of the chosen IS distribution and one that is used as a safeguard. 
In practice, the distribution used as the safeguard is usually the original distribution. 
The idea was further extended and improved in \cite{owen2000safe}. 

In this work, we provide an alternative approach to alleviate the issue. 
The proposed method is based upon the assumption that we have a ``reasonably good'' IS distribution, in the sense that, 
the chosen IS distribution is appropriate (namely, can \emph{reduce} the estimator variance) in the 
region that has dominant contribution to the integral (in what follows we shall refer to such a region as a 
``safe'' region), and the region in which the IS distribution may possibly cause problem, i.e., resulting in unbounded weight function as is explained in Section~2, has relatively small contribution to the integral. 
A more detailed explanation of the assumption can be found in Section~\ref{sec:wbis}.
Under this assumption, the implementation of the method is actually quite straight forward: given an IS distribution, we write the sought integral as the sum of two parts: one is integrated  over the ``safe'' region and one over its compliment;
based on our assumption, the integral in the ``safe'' region contributes dominantly to the total integral value, 
we can simply use that as an approximation to the total integral value and apply IS to estimate it.
As we know that IS is good in the safe region, we will obtain an estimate with high accuracy. 
As such,  we obtain an IS estimator which is biased but guaranteed  to have a  finite variance. 
A key issue in this idea is how to identify the safe region, 
and as will be discussed in Section~\ref{sec:wbis}, we define the safe region as the region 
in which the weight function is bounded by a prescribed threshold value, which insure that the IS estimator has a finite variance in the region. 
We then present a normality test based method to compute a suitable threshold value from the samples. 
{  In most practical problems, it is usually difficult to know in advance whether the IS distribution in use may cause problem, 
and the proposed method can automatically determine it and adjust accordingly.}
With numerical examples, we demonstrate that the proposed approach performs significantly better than the defensive IS method. 


The rest of the paper is organized as follows. In Section~\ref{sec:is} we present the standard IS and analyze that the method may result in
infinite estimator variance, and we then discuss the DIS method that was developed to address the issue in Section~\ref{sec:dis}.
In Section \ref{sec:wbis} we present in details our weight-bounded IS method. 

\section{Basics of Importance Sampling}\label{sec:is}
In this section we shall briefly introduce the method of IS to reduce the variance of the MC estimation. 
In particular we concentrate on the problem of computing the integral, 
\begin{equation}I=\int_{\mathcal{D}}f(x)p(x)\emph{d}x,\label{e:I}\end{equation}
where $p(x)$ is the probability density function of $x$ and $\mathcal{D}$ is the domain of $x$. 
 In what follows we shall refer to $p(x)$ as the nominal distribution, and when not causing ambiguity,  we shall omit
 the domain $\mathcal{D}$ in the integration.  
 Moreover, for simplicity we assume that function $f(x)$ is non-negative and is also bounded from above in the entire domain $\mathcal{D}$. 
 A practical example of such an assumption is  the failure probability estimation where $f(x)$ is a failure indicator function: $f(x)=1$ for $x\in F$ and $f(x)=0$ otherwise, where 
 $F$ is the region corresponding to system failures. 
 In practice, such an integral is often computed with a Monte Carlo estimation:
 \begin{equation}
 \hat{I}_\mathrm{MC} = \frac1n\sum_{i=1}^n f(\-X_i), 
 \end{equation}
 where $\{X_i\}_{i=1}^n$ are drawn from the distribution $p(x)$. 
 It is well known that the MC estimator $\hat{I}$ is an unbiased estimator of $I$ and its variance  is 
 \begin{equation}
\sigma^2_{MC} =\mathrm{VAR}[\hat{I}]= \frac{\mathrm{Var}[f]}{n}.
 \end{equation}
 In many practical problems, the variance of $f$ can be large and as a result, a rather large number of samples are needed to 
 obtain a reliable estimate of the integral $I$. 
 In this case, the technique of Importance Sampling (IS) can be used to improve the sampling efficiency. 
 The basic idea of the importance sampling is quite straightforward: instead of sampling from the nominal distribution, 
 we draw samples from an alternative distribution, referred to as the IS distribution in this paper,
 and then an appropriate weight is assigned to each sample so that it results in an unbiased estimator of $I$. 
 Specifically, given an IS distribution $q(\-x)$,   
 the integration in Eq.~\eqref{e:I} can be rewritten as
\begin{equation}\label{rewriteI}
I=\int_{\mathcal{D}}f(x)W(x)q(x)\emph{d}x.
\end{equation}
where the weight function 
\begin{equation} W(x) = p(x)/q(x)\label{e:weight}
\end{equation}
 is the ratio of the nominal density and the IS density. 
 Applying a standard MC estimation to Eq,~\eqref{e:I} yields the IS estimator:
 \begin{equation}\label{eq:ime}
{\hat{f}_{q}}=\frac{1}{n}\sum_{i=1}^{n}{f(X_{i})w(X_{i})},
\end{equation}
 where samples $\{X_i\}_{i=1}^n$ are drawn from the IS distribution $q(x)$. 
 It is easy to verify that the IS estimator in Eq.~\eqref{eq:ime} is also an unbiased estimator of $I$ and moreover,
 its variance is 
 \begin{equation}
 \sigma^2_{IS}=\mathrm{Var}[{\hat{f}_q}]=\frac{1}{n}(\int {f^{2}(x)w(x)}{p(x)}\emph{d}x-I^2). \label{e:varis}
 \end{equation}
   One can reduce the variance of the IS estimator by choosing an appropriate IS distribution $q(x)$. 
 It should be noted here that, to apply IS estimation, we must {choose} the IS distribution $q(x)$ such that $q(x)>0$ for any $x$ satisfying $p(x)>0$, i.e.,
 { the support of $p(x)$ is a subset of that of $q(x)$.}  

The performance of the IS estimation critically depends on the choice of the IS distribution. In fact,  if we choose 
\[q(x) =\frac{f(x)p(x)}I,\] known as the optimal IS distribution, the resulting estimator variance is zero. 
On the other hand, however, if the IS distribution is not chosen correctly, the IS estimation may suffer from excessively large variance and in some cases it may even fail. 
In particular, as can be seen from Eq.~\eqref{e:varis}, we may have trouble if $q(x)\ll p(x)$ in certain region in $\mathcal{D}$,
as in this case the variance can be arbitrary large as the weight function $w(x) = p(x)/q(x)$ can be unbounded in the domain $\mathcal{D}$. 
{ We refer to Section 2.2 in \cite{owen2000safe} for more discussions and an example of the issue.}

\section{Defensive Importance Sampling}\label{sec:dis}
 To address the issue in the standard IS method,  a method termed as  the defensive IS (DIS) was proposed in \cite{hesterberg1995weighted}.
 The basic idea of the DIS method is to construct a new IS distribution which is a mixture of  the original IS distribution and a heavy-tailed safe-guard distribution (which can often be the nominal distribution).
 Namely, if  $q(x)$ is the chosen IS density and $p(x)$ is the nominal density, the new DIS density is of the form \[q_{\alpha}(x)=\alpha p(x)+(1-\alpha)q(x),\]
where $0<\alpha<1$ is the parameter controlling the relative weight between $q(x)$ and $p(x)$. The defensive mixture sampling estimate can be written as
\[{\hat{f}_\mathrm{DIS}}=\frac{1}{n}\sum_{i=1}^{n}f(X_i)W_{\alpha}(X_{i}),\]
where $X_{i}$ are the random samples from the defensive mixture distribution $q_\alpha$. 
Unlike the standard IS which may suffer from unbounded weight function, the weight function in the DIS method is 
bounded from above: 
\[W_{\alpha}(x)=\frac{p(x)}{q_{\alpha}(x)}=\frac{p(x)}{\alpha p(x)+(1-\alpha)q(x)}\leq\frac{p(x)}{\alpha p(x)}=\frac{1}{\alpha}.\]
Now recall that that the integrand $f(x)$ is bounded above and specifically we assume $f(x)\leq M$ for a positive constant $M$.
It {follows} directly that the variance of the DIS estimator is no {greater than}:  
\begin{equation}
\sigma_\mathrm{DIS}^{2}=\mathrm{VAR}[{\hat{f}_\mathrm{DIS}} ]
\leq \frac{1}{\alpha}\sigma^2_\mathrm{MC}+(\frac{1}{\alpha}-1)I^2. \label{eq:var}
\end{equation}
That is, unlike the standard IS, the DIS estimator is guaranteed to have  a bounded variance (recall that $0<\alpha<1$). 
From Eq.~\eqref{eq:var}, one can see that the performance of DIS depends critically on the choice of $\alpha$. 
One can see that the upper bound in Eq.~\eqref{eq:var} is minimized at $\alpha=1$, which implies that
  if we take $\alpha\rightarrow1$, the upper bound in Eq.~\eqref{eq:var} becomes smaller;
  however, taking $\alpha \rightarrow1$ also implies that the estimator becomes close to the standard MC estimation,
  which may result very large variance, especially in the case where the IS distribution is very effective.
  To address the problem we shall provide an alternative method to prevent unbounded variance in the next section.  

\section{Weight-bounded Importance Sampling}\label{sec:wbis}

 First we choose a positive number {$r>0$ and rewrite $\E[f]$ }as,  
\begin{equation}
\E[f] = \E_r[f]+\E_{\bar{r}}[f],
\end{equation}        
where          
\[\E_r[f] = \int_{\{\-x| W(\-x)\leq r\}}  f(\-x)W(\-x) q(\-x) d\-x = \E_q[fWI_r  ] ,    \]
\[\E_{\bar{r}}[f] = \int_{\{\-x| W(\-x)> r\}}  f(\-x) W(\-x) q(\-x) d\-x = \E_q[fWI_{\bar{r}} ], \] 
and $I_r(\-x)$ and $I_{\bar{r}}(\-x)$ are two indicator functions:
\begin{equation*}
I_r(x)=
\begin{cases} 0&W(x)>r\\
1& W(x)\leq r
\end{cases},
\quad
I_{\bar{r}}(x)=
\begin{cases} 1&W(x)>r\\
0& W(x)\leq r
\end{cases}.
\end{equation*}
Now suppose we use the approximation: $\E[f]  \approx\E_r[f]$,  estimated as
\begin{equation}
\hat{f}_\mathrm{r} = \frac1n \sum_{i=1}^n f(\-x_i) W_r(\-x_i),\label{e:bis}
\end{equation}
where the samples are drawn from distribution $q$,
and 
\begin{equation*}
W_r(x)=
\begin{cases} 0&W(x)>r\\
W(x)& W(x)\leq r
\end{cases}.
\end{equation*}
 { Eq.~\eqref{e:bis} is the proposed bounded-weight importance sampling estimator.}
Simply put, when the weight function of a given sample exceeds a given threshold value, 
we simply let it to be zero. Moreover, it should be clear that $\hat{f}_\mathrm{r}$ is a \emph{biased estimator} of $E[f]$, whose 
mean square error (MSE) is
\[
\mathrm{MSE}[\hat{f}_\mathrm{r}] = \mathrm{Var}[\hat{f}_\mathrm{r}]+ (\E_r[f]- \E[f])^2.
\]
Now noting that
$\mathrm{Var}[\hat{f}_\mathrm{r}] \leq  r^2 \mathrm{Var}[f] /n$, we can see that
the MSE of the WBIS estimator $\hat{f}_r$ is bounded from above.   
It is also easy to see that the following equation holds {  as long as one can take $r$ to be $\infty$}:
\[\min_{r>0} \mathrm{MSE}[\hat{f}_r]\leq\mathrm{MSE} \hat{f}_{q},\] 
which implies that if we make a good choice of $r$ (including the choice to let $r=\infty$), the weight bounded IS estimator can be at least as good as the standard IS.

A key issue in the WBIS method is to determine the weight upper bound $r$. 
 In practice, however, 
depending on the shape of the nominal density $p(x)$, the function $f(x)$ and the sampling density $q(x)$, 
and so no generally applicable value for the parameter and it has to be determined based on the specific problem. 
Ideally for a given problem, one wants to determine the upper bound in advance (namely it should not depend on the samples); this, however,
is extremely difficult as we may not have any knowledge of the problem before drawing the samples. 
In what follows we will provide a method to determine the upper bound based on the samples drawn from the IS distribution. 
The basic idea of the method is that the chosen upper bound should ensure 
that the resulting WBIS estimator $\hat{f}_r$ is of finite variance. A sufficient condition for that is 
\[\mathrm{Var}_q[ W_r(x)] 
= \int W_r^2(x) q(x)dx - { (\int W_r(x) q(x) dx)^2} <\infty.\]
Now suppose that $X_{1},...,X_{n}$ are  $n$ i.i.d samples drawn from the density $q(x)$, by the central limit theorem, if $W_r(x)$ has finite mean $\mu_{W}$ and finite variance 
$\sigma^2_{W}$, as $n$ approaches infinity,
we have, 
\[\sqrt{n}((\frac{1}{n}\sum_{i=1}^{n}W_r(X_{i}))-\mu_{W})\stackrel{d}{\rightarrow}N(0,\sigma^{2}_{W}),\]
or equivalently  \[1/n\sum_{i=1}^{n}W_r(X_{i})\stackrel{d}{\rightarrow}N(\sqrt{n}\mu_{W},\sigma^{2}_{W}).\]
{ Thus if the variance of $W_r(x)$ is finite  
$1/n\sum_{i=1}^{n}W_r(X_{i})$} is normally distributed for sufficiently large sample size $n$.  
We shall use this to design our criterion to determine $r$. 
Specifically, we divide the samples $\{X_{1},...,X_n\}$  into $n_{group}$ groups, and each group has $n_{sample}$ samples, i.e., $n_{group}n_{sample}=n$. We modify the notation a
bit and use $X_{i,j}$ to represent the $i$-th sample in the $j$-th group. Then we compute the group statistics, 
\[\bar{W}_{j}(r)=\frac{1}{n_{sample}}\sum_{i=1}^{n_{sample}}W_r(X_{i,j}),
\quad j=1,...n_{group}.\]
It should be clear that $\bar{W}_j$ depends on the value of $r$ and so here we use {$\bar{W}_j(r)$} to emphasize such a dependence. 
Now we shall choose the maximum value of $r$ subject to the condition that
 $\bar{W}_{j}(r), j=1,...,n_{group}$ can pass a normality test ( in this work we use the Anderson-Darling test~\cite{anderson1952asymptotic}, but our method does not 
 depend on any specific normality test; for a detailed comparison of normality tests, see \cite{yazici2007comparison}) with a chosen significance level. 
 An issue here is to determine the number of groups $n_{group}$ and the number of the samples in each group $n_{sample}$. 
 Roughly speaking, if we choose  larger $n_{sample}$, { we will have more reliable estimates of $\bar{W}_{j}(r)$ in each group, but on the other hand, we will have less accurate normality test
 due to the limited number of groups;
 if we use large $n_{group}$, we will have more groups but each $\bar{W}_{j}(r)$ may not be accurately estimated.}    
 While noting that the choices of the two numbers may be problem dependent, we here use choose $n_{group}=C\sqrt{n}$, then $n_{sample}=\frac{1}{C}\sqrt{n}$
 for a prescribed constant $C$ which is used to balance accuracy of the normality test and the estimation of $\bar{W}_{j}(r)$ in each group.
It is easy to see that, by choosing the two numbers this way,  as as the total number $n$ tends to $+\infty$, both $n_{group}$ and $n_{sample}$ tend to $+\infty$. In next section we demonstrate that the proposed method performs well in several examples.

\section{Numerical examples}
\subsection{A mathematical example}

Our first example is one used in \cite{hesterberg1995weighted} to demonstrate the failure of standard IS, with slight modification. 
 Let $\mathcal{D}=(-0.5,0.5)^{5}$ and the nominal distribution be a uniform distribution: $p(x)=U(-0.5,0.5)^{5}$. The integrand is
 \begin{equation}
 f(x)=0.8\prod^{5}_{j=1}\mathcal{N}_{mul}(x^{j},2)+0.2\prod_{j=1}^{5}\{\mathcal{N}_{mul}(x^{j},2)+10^{-3}-2\times10^{-3}I_{[-\frac{1}{4},\frac{1}{4}]}(x^j)\}
 \end{equation}
 where $I_B(x^j)$ is the indicator function for region $B$, and 
 \[\mathcal{N}_{mul}(x,\theta)=\beta(\theta)(\varphi(\theta x)-\varphi(\frac{1}{2}\theta )),\quad \beta(\theta)=\frac{1}{(\frac{\Phi(\frac{1}{2}\theta)-\Phi(-\frac{1}{2}\theta)}{\theta}-\varphi(0.5))},\]
 with  $\varphi(x)$ and $\Phi(x)$ being the probability density function and the cumulative distribution function of the standard normal distribution respectively. The optimal distribution is {  $f(x)p(x)/I=f(x)p(x)$ as we note that $I=1$ in this example.}
We {choose} the IS distribution to be $q(x)=\prod^{5}_{j=1}\mathcal{N}_{mul}(x^{j},2)$.
In Figure 1 (left), we plot the IS distribution $q(x)$ and the optimal distribution $f(x)p(x)$ for the first dimension (all the dimensions are the same). 
In Fig.~1 (a) we can see that the IS distribution $q$ and $f$ agree quite well in their main lobes;
however, the sampling density $q$ tends to zero moving away from the mean,  { while by design the function $f(x)$ bounded below by a positive constant $10^{-3}$}. 
It can be verified that the variance if the IS estimator is unbounded, i.e., $Var(\hat{I}_{q})=+\infty$, and thus the problem poses a challenge to standard IS simulation. 
 \begin{figure}
\centerline{\includegraphics[width=0.5\textwidth]{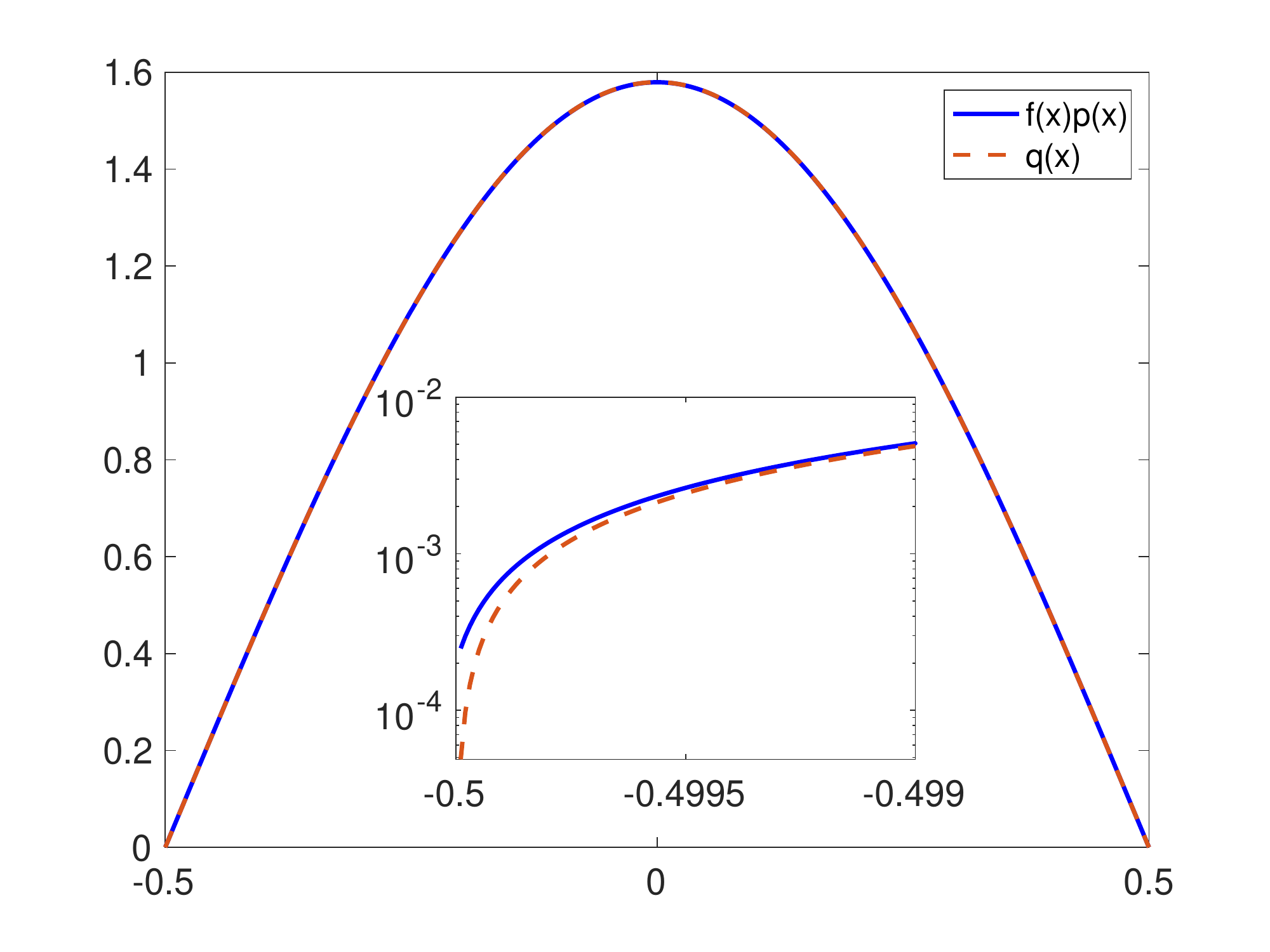}
\includegraphics[width=0.5\textwidth]{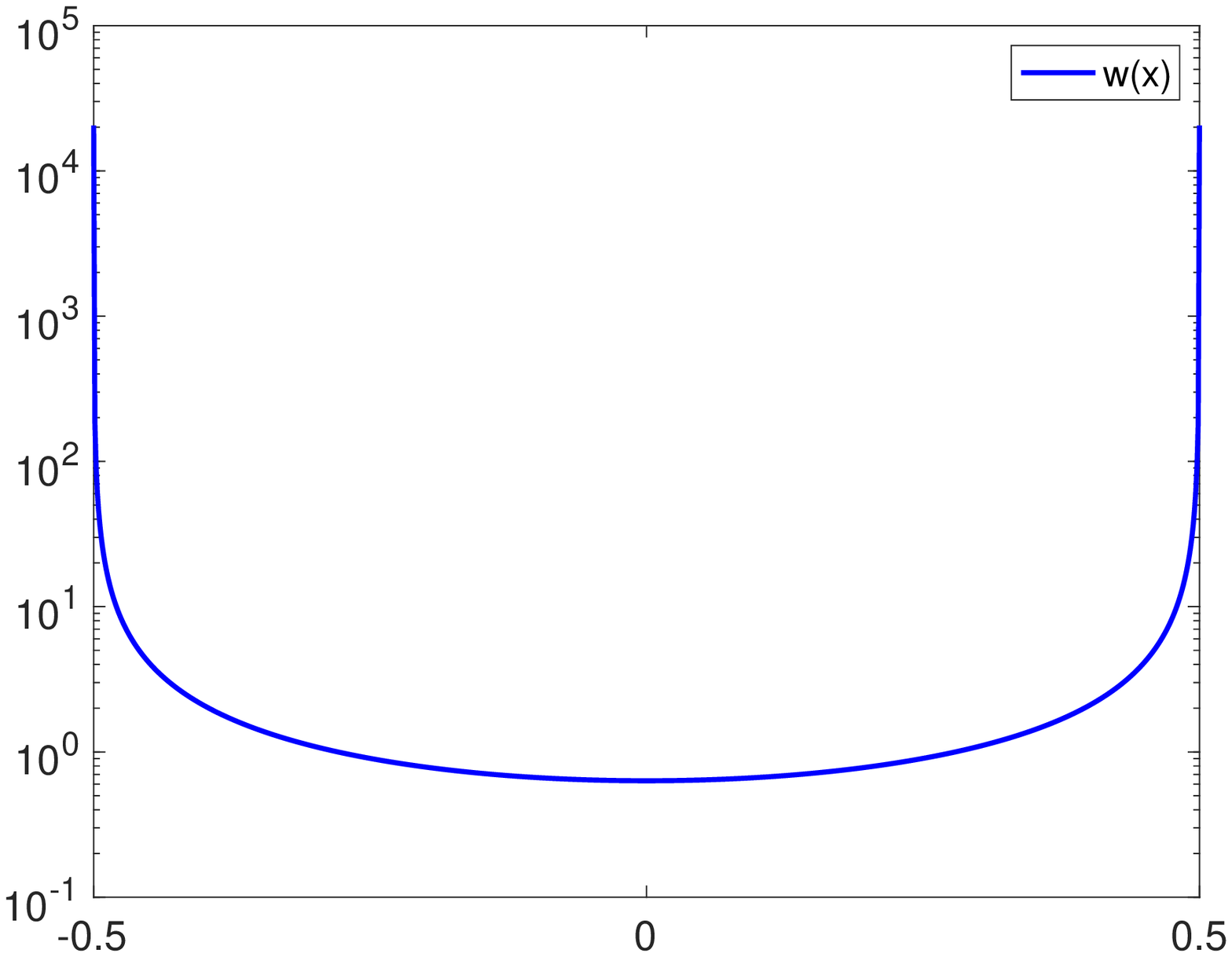}}
  \caption{Left: a comparison of the optimal distribution $f(x)p(x)$ and the chosen IS distribution $p(x)$; inset is the zoom-in plot around the tail $-0.5$ on a logarithmic scale.
  Right: the weight function.}
  \label{fig1:subfig} 
 \end{figure}

We estimate $\E[f]$ with three different methods: standard IS, DIS, and the proposed WBIS, all with the chosen IS distribution $q$. 
In the DIS method, we use { two} different values of $\alpha$: $\alpha=0.1$, $\alpha=0.5$;
in the WBIS method, we use two different significant levels: $5\%$ and $1\%$. 
For each methods we compute the estimates of $I$ with 4 different sample size: $10^4$, $4\times10^4$, $9\times10^4$ and $16\times10^4$ and for each sample size, we repeat the simulation for $10^5$ times. 
To characterize the performance of each method, we compute the {normalized} mean square error~(NMSE), 
\begin{equation}\label{NMSE}
NMSE=\frac{N}{K}\sum_{k=1}^{K}(\hat{I}_{k}-I)^{2},
\end{equation}
where $\hat{I}_{k}$ is the $k$-th estimate of $I$, { $K=10^5$ is the number of simulations performed
and $N$ is the sample size used in each simulation.}
We summarize the simulation results in Table~\ref{table:1}. { 
Also shown in Table~1 is the values of $r$ computed by our method.}
As we can see from the table, the NMSE of the standard IS increases with respect to sample size, 
and this is actually unsurprised as the variance of IS is infinity. 
On the other hand, the NMSE of the DIS is well bounded and does not vary much with respect to the sample size, 
which indicates that the DIS estimator has a finite variance. 
However, one can see here that the NMSE of DIS with $\alpha=0.5$ is about 10 times of that with $\alpha=0.1$, 
suggesting that the performance of the method is very sensitive to the choice of $\alpha$.
The table shows that, just like the DIS method, the NMSE of the proposed WBIS method remains about the same level as 
the sample size increases, and more importantly the NMSE values of WBIS results are much smaller than that of the DIS method with both significance levels, 
demonstrating a substantially better performance than DIS. To further analyze the WBIS estimator, we list
the bias (squared) and the variance in Table~\ref{table:biasvar} for significance levels $1\%$ and $5\%$.
We can see that, in all the results, the bias in the estimator is smaller than the variance of it.

\begin{table}[!htbp]

\begin{tabular}{|c|ccccc|}
  \hline
   \backslashbox{sample size}{method} &{IS}&{DIS ($\alpha=0.1$)}&{DIS ($\alpha=0.5$)}&{WBIS ($5\%$)}&WBIS ($1\%$)\\
  \hline
   \begin{minipage}{1em} \vspace{0.2cm}\centering 10000 \vspace{0.3cm}\end{minipage}&\makebox[3em]{0.144}&\makebox[3em]{0.0281}&\makebox[3em]{0.320}&\makebox[3em]{$1.479\times10^{-4}$}&$4.070\times10^{-5}$\\
  \hline
   \begin{minipage}{1em} \vspace{0.2cm}\centering 40000 \vspace{0.3cm}\end{minipage}&\makebox[3em]{1.039}&\makebox[3em]{0.034}&\makebox[1em]{0.325}&\makebox[3em]{$1.825\times10^{-4}$}&$4.865\times10^{-5}$\\
  \hline
   \begin{minipage}{1em} \vspace{0.2cm}\centering 90000 \vspace{0.3cm}\end{minipage}&\makebox[3em]{3.820}&\makebox[3em]{0.040}&\makebox[1em]{0.325}&\makebox[3em]{$2.718\times10^{-4}$}&$6.123\times10^{-5}$\\
  \hline
   \begin{minipage}{1em} \vspace{0.2cm}\centering 160000 \vspace{0.3cm}\end{minipage}&\makebox[3em]{8.628}&\makebox[3em]{0.049}&\makebox[1em]{0.330}&\makebox[3em]{$2.928\times10^{-4}$}&$6.619\times10^{-5}$\\
  \hline
\end{tabular}
\caption{The NMSE of the three methods with different sample sizes.}\label{table:1}
\end{table}

\begin{table}[!htbp]
 \centering
 \begin{tabular}{|c|cccccc|}
  \hline
  {sample size} &{Bias$^2$($5\%$)}&{Var($5\%$)}&$r$($5\%$)&{Bias$^2$($1\%$)}&{Var($1\%$)}&$r$($1\%$)\\
  \hline
  \begin{minipage}{1em} \vspace{0.2cm}\centering 10000 \vspace{0.3cm}\end{minipage}&\makebox[6em]{$2.50\times10^{-9}$}&\makebox[6em]{$1.23\times10^{-8}$}&366&\makebox[6em]{$6.00\times10^{-10}$}&\makebox[6em]{$3.47\times10^{-9}$}&399\\
  \hline
  \begin{minipage}{1em} \vspace{0.2cm}\centering 40000 \vspace{0.3cm}\end{minipage}&\makebox[6em]{$1.00\times10^{-9}$}&\makebox[6em]{$3.57\times10^{-9}$}&716&\makebox[6em]{$2.82\times10^{-10}$}&\makebox[6em]{$9.35\times10^{-10}$}&844\\
        \hline
  \begin{minipage}{1em} \vspace{0.2cm}\centering 90000 \vspace{0.3cm}\end{minipage}&\makebox[6em]{$5.97\times10^{-10}$}&\makebox[6em]{$2.42\times10^{-9}$}&1087&\makebox[6em]{$1.74\times10^{-10}$}&\makebox[6em]{$5.06\times10^{-10}$}&1295\\
        \hline
  \begin{minipage}{1em} \vspace{0.2cm}\centering 160000 \vspace{0.3cm}\end{minipage}&\makebox[6em]{$4.36\times10^{-10}$}&\makebox[6em]{$1.39\times10^{-9}$}&1446&\makebox[6em]{$1.26\times10^{-10}$}&\makebox[6em]{$4.13\times10^{-10}$}&1741\\
        \hline
 \end{tabular}
 \caption{The bias (squared), the variance, and the threshold $r$ in the WBIS estimators.}\label{table:biasvar}
\end{table}

\subsection{Portfolio Credit Risk Problem}
Our next example is a real-world problem: the portfolio credit risk problem studied in \cite{glasserman2005importance}. 
Previous studies have mainly focused on how to obtain a good IS distribution for this model. 
Here we shall apply our WBIS method to provide a ``safe''  { estimate of the default probability. In this problem, we consider a financial institute with $m$ obligors and assess the risk of excessive losses.
The settings of the problem are shown below:}
\begin{itemize}
\item $Y_{k}$: default indicator for $k$-th obligor;
 $Y_k =1$ if  the $k$-th obligor defaults, $Y_k=0$ otherwise; 
\item $p_{k}$:  the  probability that the $k$-th obligor defaults;
\item $c_{k}$: the loss resulting from the default of the $k$-th obligor;
\item  $L= c_{1}Y_{1}+...+c_{m}Y_{m}$: the total loss from  all obligors.
\end{itemize}
We take the individual default probabilities $p_{k}$ and the loss $c_{k}$ as constants for simplicity, and  the goal is to estimate the default probability 
 $P=\mathbb{P}(L>x)$ for a prescribed loss threshed $x$.
Next we shall describe how the default of an obligor is defined. 
We characterize the default indicator $Y_{k}$  by the vector$(X_{1},...,X_{m})$ of latent variables. Specifically $Y_k$ is given by, 
\[Y_{k}=\pmb{I}_{\{X_{k}>x_{k}\}},\ \  k=1,...,m\]
with $x_{k}$ chosen to match the marginal default probability $p_{k}$. 
Moreover, the latent variables $X_k$ are assumed to have the form of
\[X_{k}=a_{k1}Z_{1}+...+a_{kd}Z_{d}+b_{k}\epsilon_{k},\ \ k=1,...m,\]
in which
\begin{itemize}
           \item $Z_{1},...Z_{d}$ are systematic risk factors, each having an independent standard normal distribution;
           \item $\epsilon_{k}$ is an idiosyncratic risk associated with the $k$-th obligor, each following an independent 
           standard normal distribution;
           \item $a_{k1},...a_{kd}$ are the factor loadings for the $k$-th obligor, $a_{k1}^{2}+...+a_{kd}^2\leq 1$;
           \item $b_{k}=\sqrt{1-(a_{k1}^2+...+a_{kd}^2)}$.
\end{itemize}
In the example, the portfolio has 10 systematic risk factors, and there are $m=1000$ obligors in the market. The other settings are
\begin{align*}
&p_{k}=0.01(1+\sin(16\pi k/m)),\ \ k=1,...,m;\\
&c_{k}=(\left \lceil 5k/m \right \rceil)^{2},\ \ k=1,...,m.
\end{align*}
Firstly, we generate the a group of parameters $a_{k1},...,a_{kd}$ and $b_{k}$ for $k=1,...,m$ from a unit ball satisfy $(a_{k1}^2+...+a_{kd}^2)+b_{k}^2=1$ .
We then choose the threshold loss value to be $x=9500$, and by a direct MC simulation with $10^9$ samples, we estimate that the 
default probability is $3.5\times10^{-6}$, which is regarded as the actual value of the default probability.  
We assume the IS distribution of Gaussian with its mean and covariance determined by using the cross-entropy method~\cite{deBoerEtAl_AOR05,RubinsteinK_04}. 
{  In the cross-entropy method, we use diagonal covariance matrix; moreover, as the specific mean and covariance
estimates are highly problem dependent, we choose to omit them here.}
We use IS, DIS and WBIS to  estimate the default probability $P$.  
{  We emphasize here that, direct use of the IS method may potentially result in an unbounded variance,
while DIS and WBIS can provide a ''safe'' estimation of the sought probability.} 

Specifically, for the  DIS method we use $\alpha=0.1$ and $\alpha=0.5$, and for our WBIS method
we use the same two significance  levels as is in the first example: $1\%$ and $5\%$. To obtain a reliable comparison, with each method we estimate $P$ using $10^4$ samples and repeat
the simulations $2000$ times. We then can compute the root mean square error (RMSE) of the 2000 estimates for either method:{  
\begin{equation}
\mathrm{RMSE} = \sqrt{\frac1M \sum_{i=1}^{M} (P_i-P)^2},
\end{equation}
where $P$ is the exact value of the sought probability, $P_i$ is the $i$-th estimate of $P$,} and $M$ is the total number of estimates, which in this example is 2000. 
We summarize the RMSE results in Table~\ref{table:3}.

\begin{table}
\centering
\begin{tabular}{|c|ccccc|}
\hline
  Method&IS&DIS($\alpha=0.1$)&DIS($\alpha=0.5$)&WBIS ($1\%$)&WBIS($5\%$)\\
 &&&&&\\
  \hline
 &&&&& \\
  { RMSE}&$3.2\times10^{-6}$&$6.2\times10^{-6}$&$6.8\times10^{-6}$&$2.9\times10^{-6}$&$3.3\times10^{-6}$\\
  \hline
\end{tabular}
\caption{The { RMSE} of the three methods with different parameter values.}\label{table:3}
\end{table}

From the table we can see that, the RMSE of the proposed WBIS method is evidently lower than that of the DIS method, regardless of what parameter values are used. Moreover, our numerical results also suggest that the WBIS method is not  sensitive to the choice of the significance level, and in practice it is reasonable to use either $1\%$ or $5\%$. 
It should be noted here that the IS method also achieves rather good accuracy, suggesting that 
in this example, the chosen IS distribution actually performs well, 
and in this case, the proposed WBIS method produces comparable results, while DIS significantly increases the variance. 

\section{Conclusions}
In this paper, we consider the problems where standard IS simulation may have the risk of unbounded variance and we propose
a weight bounded IS method to address the issue. 
The method assumes that the IS distribution is appropriate in the region that has dominant contribution to the integral, i.e., the safe region,
and the method  performs a standard IS in this safe region and use the resulting estimate as an approximation to the original integral. 
We then propose a normality test based method to identify the safe region from samples. 
With numerical examples we demonstrate that the proposed method can result in bounded estimator variance when standard IS fails,
and more importantly it can yield more accurate estimates than the often used defensive IS method. 
In summary, we believe that the proposed WBIS method can be useful in a large class of problems where standard IS simulation may become problematic~(i.e., 
resulting in unbounded variance). 
We plan to investigate the application of the WBIS method to some real world problems of this type in the future.  
\section*{Acknowledgements}
L. Lu is supported by the National Natural Science Foundation of China under grant number 5150080805. J. Li is supported by the National Natural Science Foundation of China under grant number 11771289. 

\bibliography{wbis}{}
\bibliographystyle{plain}
\end{document}